\documentclass[a4paper,12pt,useAMS]{mn2e}

\usepackage{graphicx}

\title[Strong-Lensing Analysis of MS 1358.4+6245]{Strong-Lensing Analysis of MS 1358.4+6245: New Multiple Images and Implications for the Well-Resolved z=4.92 Galaxy}
\author[Zitrin et al.]{Adi Zitrin$^{1}$\thanks{E-mail:
adiz@wise.tau.ac.il}, Tom Broadhurst$^{2,3}$, Dan Coe$^{4}$, Jori Liesenborgs$^{5}$, Narciso Ben\'itez$^{6}$, \and Yoel Rephaeli$^{1}$, Holland Ford$^{7}$, Keiichi Umetsu$^{8}$\\\\\\
$^{1}$The School of Physics and Astronomy, the Raymond and Beverly Sackler Faculty of Exact Sciences, Tel Aviv University,\\ Tel Aviv 69978, Israel\\
$^{2}$Department of Theoretical Physics, University of Basque Country UPV/EHU, Leioa, Spain\\
$^{3}$IKERBASQUE, Basque Foundation for Science, 48011, Bilbao, Spain\\
$^{4}$Jet Propulsion Laboratory, California Institute of Technology, 4800 Oak Grove Dr, MS 169-327, Pasadena, CA 91109\\
$^{5}$Expertisecentrum voor Digitale Media, Universiteit Hasselt, Wetenschapspark 2, B-3590, Diepenbeek, Belgium\\
$^{6}$Instituto de Astrof\'isica de Andaluc\'ia (CSIC), C/Camino Bajo de Hu\'etor, 24, Granada, 18008, Spain\\
$^{7}$Department of Physics and Astronomy, Johns Hopkins University, 3400 North Charles Street, Baltimore, MD 21218\\
$^{8}$Institute of Astronomy and Astrophysics, Academia Sinica, P.~O. Box 23-141, Taipei 10617, Taiwan}
\begin{document}


\pagerange{\pageref{firstpage}--\pageref{lastpage}} \pubyear{2010}

\maketitle

\label{firstpage}

\begin{abstract}

We present a strong-lensing analysis of the galaxy cluster MS
1358.4+6245 ($z=0.33$), in deep 6-band ACS/HST imaging. In addition to
the well-studied system at $z=4.92$, our modelling method uncovers 19
new multiply-lensed images so that a total of 23 images and their
redshifts are used to accurately constrain the inner mass
distribution. We derive a relatively shallow inner mass profile,
$d\log \Sigma/d\log r\simeq -0.33 \pm0.05$ ($r<200$ kpc), with a much higher
magnification than estimated previously by models constrained only by the $z=4.92$ system. Using these many new images we can apply a
non-parametric adaptive-grid method, which also yields a shallow mass
profile without prior assumptions, strengthening our conclusions. The
total magnification of the $z_s=4.92$ galaxy is high, about a
$\sim100\times$ over its four images, so that the inferred source
size, luminosity and star-formation rate are about $\sim5\times$
smaller than previous estimates, corresponding to a dwarf-sized galaxy
of radius $\simeq1$ kpc. A detailed image of the interior morphology
of the source is generated with a high effective resolution of only $\simeq$50 pc,
thanks to the high magnification and to the declining angular diameter
distance above $z\sim1.5$ for the standard cosmology, so that this image apparently represents the best resolved object known at high redshift.


\end{abstract}


\section{Introduction}

The galaxy cluster MS 1358.4+6245 (also known as CL 1358+62; hereafter
MS 1358) was first identified by Zwicky \& Herzog (1968), and later
classified as an X-ray luminous, rich cluster by Luppino et
al. (1991) based on observations from the
\emph{Einstein Observatory Extended Medium-Sensitivity Survey} (EMSS;
Gioia et al. 1990, Stocke et al. 1991). This cluster has been subject
to intensive study, mainly due to its high X-ray luminosity and
richness at an intermediate-high redshift (e.g., Henry et al. 1992,
Carlberg et al. 1996, Poggianti \& Barbaro 1996, Mushotzky \& Scharf
1997, Nichol et al. 1997, Kelson et al. 1997), but became particularly
famous when a pair of $z=4.92$ lensed galaxies were uncovered in its
inner region by Franx et al. (1997). Subsequently,
other such high-$z$ galaxies have been found in systematic surveys of
galaxy clusters and in deep field imaging (e.g., Frye
\& Broadhurst 1998, Frye, Broadhurst
\& Ben\'itez 2002, Kneib et al. 2004, Stark et al. 2007, Bouwens et
al. 2004, 2009b, 2010, Bradley et al. 2008, Zheng et al. 2009).

Magnified objects are particularly useful, being generally
 bright enough to yield useful spectra, in particular the $z=4.92$
 system in MS 1358 which has an especially well-resolved large fold
 image straddling the tangential critical curve. Detailed spectroscopy
 by Franx et al. (1997; see also Swinbank et al. 2009) has revealed
 asymmetric and redward shifted Ly-alpha emission relative to metal
 absorption lines, which is claimed to imply this galaxy is suffering
 an outflow of metal enriched gas (Franx et al. 1997), similar to local
 dwarf starburst galaxies (Dekel \& Silk 1986, Heckman, Armus \& Miley
 1990). This spectral signature of gas outflow was shown to be a
 general property of $z>4$ galaxies, in the larger sample of lensed
 galaxies behind eight massive clusters, by Frye, Broadhurst \&
 Ben\'itez (2000), with implications for the enrichment of the IGM in
 general (Scannapieco \& Broadhurst 2001).

In the past two decades other extensive studies have included or
focused on MS 1358, ranging from X-ray to radio measurements and
scaling relations (e.g., Cagnoni, della Ceca \& Maccacaro 1998, Cooray
et al. 1998, Ettori \& Fabian 1999, Allen 2000, Wu 2000, B\"ohringer
et al. 2000, Arabadjis, Bautz \& Garmire 2002, Brown et al. 2003,
McCarthy et al. 2003, Egami et al. 2006, Laroque et al. 2006, Morandi,
Ettori \& Moscardini 2007), to intracluster content and evolution
studies (e.g., van Dokkum et al. 1998, Kelson et al. 2000a,b, 2006,
Borgani et al. 1999, Fabricant, Franx \& van Dokkum 2000, Ferreras \&
Silk 2000, Henry 2000, Kochanek et al. 2000, Tran et al. 2003, Postman
et al. 2005, Holden et al. 2007), and arc statistics and searches (La
F\`evre et al. 1994, Hattory, Watanabe \& Yamashita 1997, Luppino et
al. 1999, Oguri, Lee \& Suto 2003, Wambsganss, Bode
\& Ostriker 2004, Sand et al. 2005).

Despite this broad study, only one multiply-lensed system has been
hitherto identified in this cluster (the $z=4.92$ drop-out galaxy;
Franx et al. 1997), so that no full SL analysis has been possible
given the wide degeneracy of models with so few images. Franx et
al. (1997) have presented a SL mass model based on the $z=4.92$ drop-out
galaxy they uncovered, and various
simplified mass models have been used by others (e.g., Allen 1998,
Molikawa et al. 1999, Williams, Navarro \& Bartelmann 1999), often
based on isothermal potentials or spherical symmetry. Richard et al. (2008; see also Swinbank
et al. 2009) have presented two $z\sim7.5$ unverified multiply-lensed
candidates for which Ly-alpha emission has not been detected. Moreover, we note that
for a cluster at $z_{cl}=0.33$, the lensing-distance ratio for sources
at $z_s=4.92$ and $z_s\sim7.5$ differs by only $\sim2\%$, so that in order
to meaningfully constrain the inner mass profile it is crucial to use
other multiply-lensed systems at lower redshifts to expand the range of
lensing-distances.

Here we use our well-tested approach to modelling in order find a
significant number of multiple images across the central field of MS
1358 so that the mass distribution \emph{and its profile} can be well
constrained. This method was developed by Broadhurst et al. (2005a),
and simplified further by Zitrin et al. (2009b) and has securely
identified tens of multiple images in high quality HST/ACS (Advanced
Camera for Surveys) images, of background sources behind several
clusters with deep ACS/HST imaging, including Abell 1689, Cl0024+17
and a sample of 12 MACS clusters at $z>0.5$ (Broadhurst et al. 2005a, Zitrin et al. 2009b, Zitrin et
al. 2010a). This is done with only six free parameters so that in
practice the number of multiple images uncovered readily exceeds the
number of free parameters as minimally required in order to obtain a reliable fit.

This approach to lens-modelling is based on the reasonable assumption that mass approximately traces light.
Recently we have independently tested this assumption in Abell 1703
(Zitrin et al. 2010b), by applying the non-parametric technique of
Liesenborgs et al. (2006, 2007, 2009) for comparison, which is
also employed here. This latter technique employs an adaptive grid
inversion method and does not require prior assumptions regarding the mass
distribution, relying only on the images we have identified and their
redshift estimates. For meaningful constraints, this approach to
modelling requires many sets of multiple images, and over a wide range of
background redshifts per cluster. This condition is well met for A1703
(Zitrin et al. 2010b) and this model-independent method
yields a very similar mass distribution to our parametric technique in
the case of A1703, supporting the assumption that mass
generally traces light. Independently, it has been found that SL
methods based on parametric modelling are accurate at the level of a
few percent in determining the projected inner mass (Meneghetti et al. 2010).

The results of this work will be further combined with a wide-range WL
data from deep Subaru imaging to provide the cluster mass profile out
to the virial radius and beyond (Umetsu et al., in preparation). To
date only a few clusters have been reliably analysed by combining both
weak and strong lensing for a full determination of the mass profile
and a definitive comparison with theoretical predictions, (e.g.,
Gavazzi et al. 2003, Broadhurst et al. 2005b, 2008, Merten et
al. 2009, Newman et al. 2009, Okabe et al. 2009 and references
therein, Umetsu et al. 2010, Zitrin et al. 2010b). The upcoming
multi-cycle HST program of cluster imaging (the \emph{CLASH}
program$^{1}$\footnotetext[1]{PI: Postman;
http://www.stsci.edu/$\sim$postman/CLASH/}) will provide a much more
definitive derivation of mass profiles for a statistical sample of
relaxed, X-ray selected clusters, combining high resolution space
imaging with deep, wide-field ground based data, for a definitive
determination of the equilibrium mass profiles of virilised clusters.

Luppino et al. (1991) have measured the Brightest Cluster Galaxy (BCG) of
MS 1358 at a redshift of $z=0.323$. Other spectroscopic studies of
many cluster member galaxies suggest a similar cluster redshift of
$z\simeq0.33$, which is the redshift adopted here (e.g., Fabricant,
McClintock \& Bautz 1991, Carlberg et al. 1996, Yee et al. 1998,
Fisher et al. 1998).

The paper is organised as follows: In \S 2 we describe the
observations. In \S 3 we detail the SL analysis and in \S 4 we report
and discuss the results, which are then summarised in \S5. Throughout
this paper we adopt a concordance $\Lambda$CDM cosmology with
($\Omega_{\rm m0}=0.3$, $\Omega_{\Lambda 0}=0.7$, $h=0.7$). With these
parameters one arcsecond corresponds to a physical scale of 4.75 kpc
for this cluster (at $z=0.33$). The reference centre of our analysis
is fixed at the centre of the BCG: RA = 13:59:50.55 Dec = +62:31:05.00
(J2000.0).

\section{Observations and Photometric Redshifts}\label{obs}

MS 1358 was observed with the Wide Field Channel (WFC) of the ACS
installed on HST in the framework of the ACS Guaranteed Time
Observations (GTO; PI: Ford, H.; program IDs 9292, 9717, 10325) which
includes deep observations of several massive, intermediate-redshift
galaxy clusters (Ford et al. 2003).  Integration times of 7928, 5470,
5482, 9196, 13552 and 17757 seconds were obtained through the F435W,
F475W, F555W, F625W, F775W, and F850LP ($B g\arcmin V r\arcmin
i\arcmin z\arcmin$) filters, respectively.  Some important aims of the
GTO program were the determination of the mass distribution of clusters for
testing the standard cosmological model and to study distant,
background lensed galaxies for which some of the very highest redshift
galaxies are found because of high magnification by massive clusters.

Mass models and detailed lensing analyses have been presented for most
of the GTO clusters (e.g., Broadhurst et al. 2005a, El\'iasd\'ottir et al. 2007, Jee et al. 2007, Lemze et
al. 2008, 2010, Limousin et al. 2008,
Richard et al. 2009, Umetsu et al. 2010, Zitrin et al. 2009b, 2010b, Coe et al. 2010, Medezinski et al. 2010). In this work, we
present a strong lensing (SL) analysis of the ACS images of MS 1358, and
detect several multiple-image systems so the mass profile of this
cluster can be reliably constrained as well.

The ACS images were initially reduced, processed, and analysed by
APSIS, the ACS GTO pipeline (Blakeslee et al. 2003).  An optimal
$\chi^2$ detection image was created as a weighted sum of all filters,
each divided by its background RMS.  Objects were detected and
photometry obtained using SExtractor (Bertin \& Arnouts 1996).

The light of cluster galaxies was carefully modelled in each filter
and subtracted from the images.  This improves both the detection and
photometry of lensed background objects.  Based on this $B g\arcmin V
r\arcmin i\arcmin z\arcmin$ photometry, we obtain photometric
redshifts using BPZ (Ben\'itez 2000, Ben\'itez et al. 2004, Coe et
al. 2006).  The distances to the galaxies are, of course, key
ingredients to the lens model.


\section{Strong Lensing Modelling and Analysis}\label{model}


We apply our well tested approach to lens modelling, which has
previously uncovered large numbers of multiply-lensed galaxies in ACS
images of Abell 1689, Cl0024, and 12 high-$z$ MACS clusters
(respectively, Broadhurst et al. 2005a, Zitrin et al. 2009b, Zitrin \&
Broadhurst 2009, Zitrin et al. 2009a, 2010a,b). Briefly, the basic
assumption adopted is that mass approximately traces light, so that
the photometry of the red cluster member galaxies is used as the
starting point for our model. Cluster member galaxies are identified
as lying close to the cluster sequence by the photometry described in
\S \ref{obs}.

\begin{figure}
 \begin{center}
   \includegraphics[width=85mm, trim=-8mm 0mm 0mm 0mm,clip]{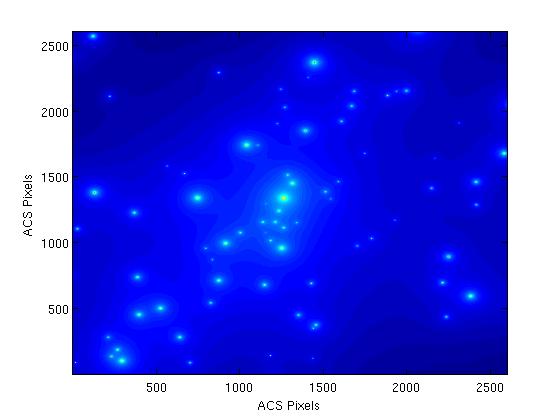}
 \end{center}
\caption{The starting point of the mass model, where we define the
surface mass distribution based on the cluster member galaxies (see \S \ref{model}). Axes are in ACS pixels ($0.05 \arcsec /pixel$).}
\label{lumpyhcomp}
\end{figure}

\begin{figure}
 \begin{center}
   \includegraphics[width=85mm, trim=-8mm 0mm 0mm 0mm,clip]{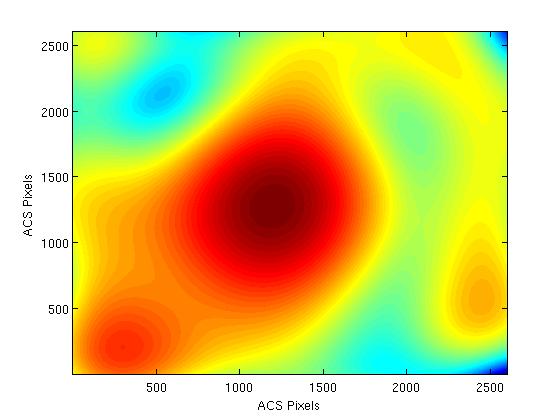}
 \end{center}
\caption{The resulting smooth mass component of the mass model (see \S \ref{model}). Axes are in ACS pixels ($0.05 \arcsec /pixel$).}
\label{smoothcomp}
\end{figure}

 We approximate the large scale distribution of cluster mass by assigning a
 power-law mass profile to each galaxy (see Figure \ref{lumpyhcomp}), the
 sum of which is then smoothed (see Figure \ref{smoothcomp}). The
 degree of smoothing ($S$) and the index of the power-law ($q$) are
 the most important free parameters determining the mass profile. A
 worthwhile improvement in fitting the location of the lensed images
 is generally found by expanding to first order the gravitational
 potential of this smooth component, equivalent to a coherent shear
 describing the overall matter ellipticity, where the direction of the
 shear and its amplitude are free parameters, allowing for some flexibility in
 the relation between the distribution of DM and the distribution of
 galaxies, which cannot be expected to trace each other in detail. The
 total deflection field $\vec\alpha_T(\vec\theta)$, consists of the
 galaxy component, $\vec{\alpha}_{gal}(\vec\theta)$, scaled by a
 factor $K_{gal}$, the cluster DM component
 $\vec\alpha_{DM}(\vec\theta)$, scaled by (1-$K_{gal}$), and the
 external shear component $\vec\alpha_{ex}(\vec\theta)$:

\begin{equation}
\label{defTotAdd}
\vec\alpha_T(\vec\theta)= K_{gal} \vec{\alpha}_{gal}(\vec\theta)
+(1-K_{gal}) \vec\alpha_{DM}(\vec\theta)
+\vec\alpha_{ex}(\vec\theta),
\end{equation}
where the deflection field at position $\vec\theta_m$
due to the external shear,
$\vec{\alpha}_{ex}(\vec\theta_m)=(\alpha_{ex,x},\alpha_{ex,y})$,
is given by:
\begin{equation}
\label{shearsx}
\alpha_{ex,x}(\vec\theta_m)
= |\gamma| \cos(2\phi_{\gamma})\Delta x_m
+ |\gamma| \sin(2\phi_{\gamma})\Delta y_m,
\end{equation}
\begin{equation}
\label{shearsy}
\alpha_{ex,y}(\vec\theta_m)
= |\gamma| \sin(2\phi_{\gamma})\Delta x_m -
  |\gamma| \cos(2\phi_{\gamma})\Delta y_m,
\end{equation}

where $(\Delta x_m,\Delta y_m)$ is the displacement vector of the
position $\vec\theta_m$ with respect to a fiducial reference position,
which we take as the lower-left pixel position $(1,1)$, and
$\phi_{\gamma}$ is the position angle of the spin-2 external
gravitational shear measured anti-clockwise from the $x$-axis.  The
normalisation of the model and the relative scaling of the smooth DM
component versus the galaxy contribution brings the total number of
free parameters in the model to 6. This approach to SL is sufficient
to accurately predict the locations and internal structure of multiple
images, since in practice the number of multiple images uncovered
readily exceeds the number of free parameters thus fully constraining
them.

\begin{figure*}
 \begin{center}
  \includegraphics[width=180mm]{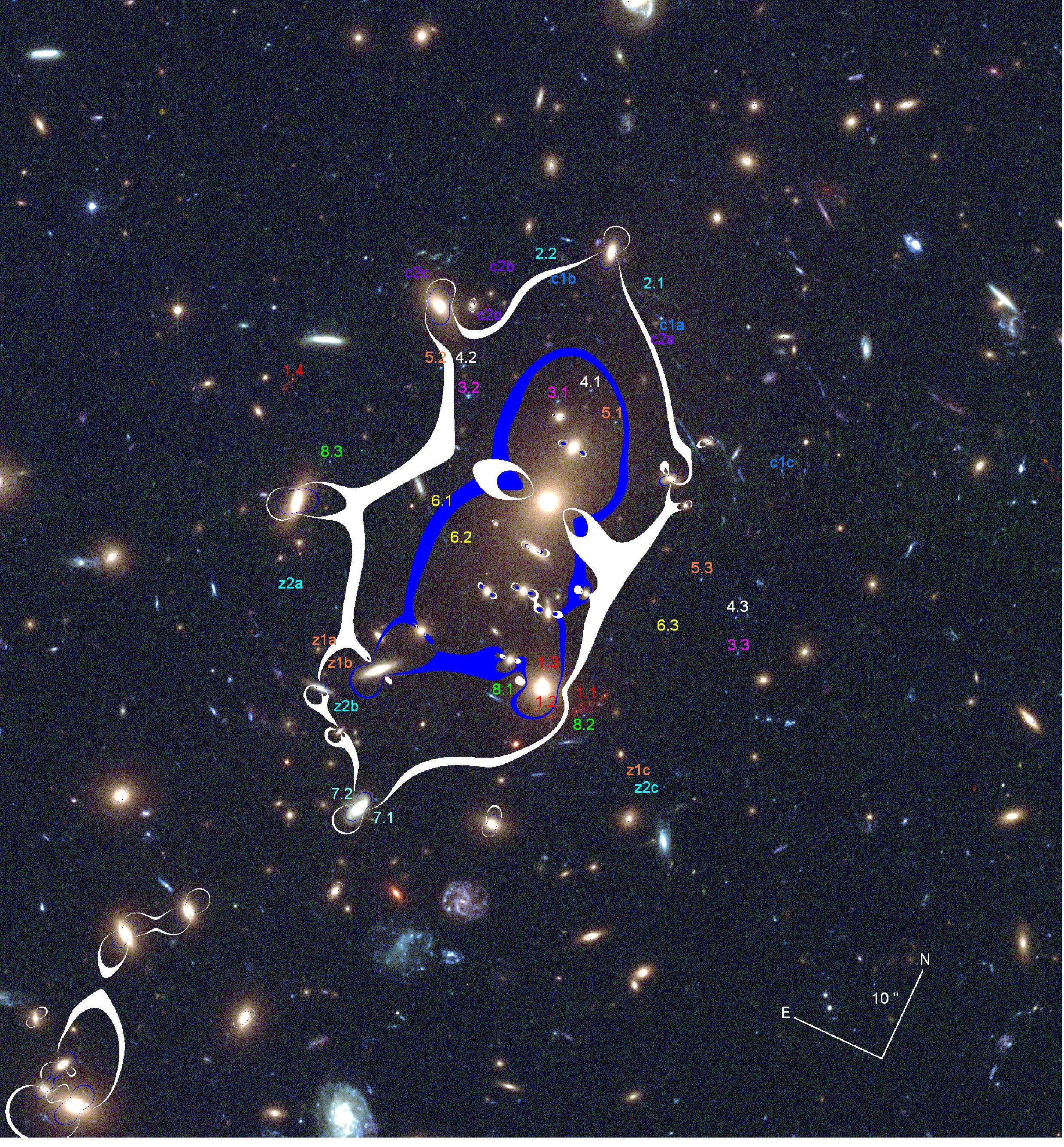}
 \end{center}
\caption{Galaxy cluster MS 1358 ($z=0.33$) imaged
with Hubble/ACS. We number the multiply-lensed images used and
uncovered in this work. The overlaid white critical curve corresponds
to system 1, the red drop-out galaxy at $z_{s}=4.92$ (found by Franx
et al. 1997), enclosing a critical area of an effective Einstein
radius of $\simeq 100$ kpc at the redshift of this cluster. Interior
to this, the blue critical curves correspond to systems 3,4,5, at a lower
redshift of $z_s\sim2$. Additional less secure candidate
systems (marked with ``c'') have similar appearance but are not
well reproduced by our best fit solution, and were not used to constrain the model. We also show the
approximate location of two unverified candidate $z_s\sim7.5$
multiple-systems, marked with ``z'' (Richard et al. 2008). These
were not used to constrain the model but their locations and claimed
high redshift are plausible in the context of our model, see \S \ref{Mimages}.}
\label{curves1358}
\end{figure*}

In addition, two of the six free parameters can be primarily set to
reasonable values so only 4 of these parameters have to be constrained
initially, which sets a very reliable starting-point using obvious
systems. The mass distribution is therefore primarily well
constrained, uncovering many multiple-images which can then be
iteratively incorporated into the model, by using their redshift
estimation and location in the image-plane.

Firstly we use this preliminary model to lens the more obvious lensed
galaxies back to the source plane by subtracting the derived
deflection field, and then relens the source plane to predict the
detailed appearance and location of additional counter images, which
may then be identified in the data by morphology, internal structure
and colour. We stress that multiple images found this way must be
accurately reproduced by our model and are not simply eyeball
``candidates'' requiring redshift verification. The best fit is
assessed by the minimum RMS uncertainty in the image plane:

\begin{equation} \label{RMS}
RMS_{images}^{2}=\sum_{i} ((x_{i}^{'}-x_{i})^2 + (y_{i}^{'}-y_{i})^2) ~/ ~N_{ima
ges},
\end{equation}
where $x_{i}^{'}$ and $y_{i}^{'}$ are the locations given by the
model, and $x_{i}$ and $y_{i}$ are the real image locations, and the
sum is over all $N_{images}$ images. The best-fit solution is unique
in this context, and the model uncertainty is determined by the
location of predicted images in the image plane. Importantly, this
image-plane minimisation does not suffer from the well known bias
involved with source plane minimisation, where solutions are biased by
minimal scatter towards shallow mass profiles with correspondingly
higher magnification.

The model is successively refined as additional sets of multiple
images are incorporated to improve the fit, importantly using also
their redshift information for better constraining the mass slope
through the cosmological relation of the $d_{ls}/d_{s}$ growth.

In order to examine our results, we also apply the
adaptive-grid, non-parametric modelling method of Liesenborgs et
al. (2006, 2007, 2009), which makes no prior assumptions regarding
the mass distribution. In a recent work, we have made a more thorough
comparison of these methods by analysing the well studied cluster
Abell 1703 (Zitrin et al. 2010b). We showed that the results of these
two distinct approaches are very similar when sufficient data constraints
are available. The effective area modelled with the non-parametric
technique is limited to roughly the critical area, beyond which
there are very few multiple-images that can be used to reliably
constrain the fit. Some discrepancy may be
expected with regards to the mass distribution between the two methods, especially when not
enough constraints are available. However, aside from minor local trade-offs
originating in the mass-sheet degeneracy or in the inclusion of cluster members, the overall profile can be expected to
be similar in order to maintain the cosmological lensing-distance
ratio of the various systems.

\section{Results and Discussion}

\subsection{Multiple-Images, Mass Model and Critical Curves}\label{Mimages}

\begin{table*}
\caption{Multiple-image systems and candidates used and uncovered by our model. For
more detailed information on each system and other optional members see the corresponding subsection. The columns are: arc ID; RA
and DEC in J2000.0; 95\% confidence level minimal and maximal
photo-$z$; best photo-$z$; $z_{model}$, the redshift predicted by the
mass model; comments. Note that system 1 was uncovered by Franx et
al. (1997) who measured their redshift spectroscopically, which is the
value given below. Note also that unusually large errors in the
photo-$z$ imply a bimodal distribution. In such cases the values which
agree with the SL model can be different than specified in the best
photo-$z$ column, as they arise from another peak in the
distribution. Such cases are specified in the comments.}
\label{systems}
\begin{center}
\begin{tabular}{|c|c|c|c|c|c|cc|}
\hline\hline
ARC & RA & DEC & Min $z_{phot}$ & Max $z_{phot}$ &Best $z_{phot}$&$z_{model}$&Comment\\
ID& (J2000.0)&(J2000.0)& (95\% C.L.)& (95\% C.L.) & & &\\
\hline
1.1 &13:59:48.689 &+62:30:48.00 &--&--&4.92& $4.92$& spec-$z$\\
1.2 &13:59:49.239 &+62:30:45.30 &--&--&4.92& "&" \\
1.3 &13:59:49.564 &+62:30:49.32 &--&--&4.92& "&" \\
1.4 &13:59:54.812 &+62:31:04.52 &--&--&4.92& "&"\\
\hline
2.1 &13:59:50.454 &+62:31:28.83 &0.25&3.93&3.42 & $3.4\pm0.3$ &Bimodal\\
2.2 &13:59:51.949 &+62:31:28.20 &0.18&3.48&0.53 & " &Bimodal\\
\hline
3.1 &13:59:51.097 &+62:31:14.73 &--&--&--& $1.9\pm0.2$ & $z_{phot}$ NA\\
3.2 &13:59:52.333 &+62:31:11.30 &1.68& 2.39& 2.04&"  &\\
3.3 &13:59:47.077 &+62:30:59.44&  1.67&  2.39& 2.03&"&\\
\hline
4.1 &13:59:50.729 &+62:31:17.28& 1.82& 2.59& 2.22& $2.1\pm0.2$& \\
4.2 &13:59:52.592 &+62:31:13.88& 1.76& 2.49& 2.13& "& \\
4.3 &13:59:47.328 &+62:31:02.60& 1.68& 2.39& 2.04& "& \\
\hline
5.1 &13:59:50.197 &+62:31:15.38&  1.67&  2.38& 2.02& $2.0\pm0.2$&\\
5.2 &13:59:52.879 &+62:31:12.70&  1.67&  2.38& 2.02& "&\\
5.3 &13:59:48.046 &+62:31:04.65 &1.67& 2.38& 2.03 & " &\\
\hline
6.1 &13:59:52.131 &+62:30:59.22&  1.61&  2.39& 2.00& $1.9\pm0.2$ &\\
6.2 &13:59:51.723 &+62:30:56.91&  1.41&  2.26& 1.77& " &\\
6.3 &13:59:48.150 &+62:30:58.05& 1.34 & 2.67 & 1.80&"&\\
\hline
7.1 &13:59:50.897 &+62:30:27.83&--&--& --& $\sim1.8$ &$z_{phot}$ NA\\
7.2 &13:59:51.330 &+62:30:28.06&--&--& --& $\sim1.8$ &$z_{phot}$ NA\\
\hline
8.1 &13:59:50.087 &+62:30:44.59& 1.76&2.50& 2.13&$2.15\pm0.2$&\\
8.2 &13:59:48.737 &+62:30:46.85& 1.84&2.60& 2.23&"&\\
8.3 &13:59:53.746 &+62:30:59.07& 1.79&2.53& 2.16& "&\\
\hline
c1a &13:59:50.252 &+62:31:27.06 & 1.77&  2.51& 2.14&$\sim3.2$&\\
c1b &13:59:51.952 &+62:31:26.86 & 1.67&  2.39& 2.03&$\sim3.2$&\\
c1c &13:59:47.656 &+62:31:18.25 & 1.68&  2.40& 2.04&---&\\
\hline
c2a &13:59:50.354 &+62:31:24.84  &  0.73 &5.53& 4.84& $\sim4-5$&Bimodal\\
c2b &13:59:52.638 &+62:31:24.04  & 0.57  & 4.81 &0.78& "&Bimodal\\
c2c &13:59:53.585 &+62:31:19.68&--&--& --& "&Similarly red\\
c2d &13:59:52.644 &+62:31:20.19&--&--& --& "&Similarly red\\
\hline\hline
\end{tabular}
\end{center}
\end{table*}

\begin{figure}
 \begin{center}
   \includegraphics[width=85mm, trim=-8mm 0mm 0mm 0mm,clip]{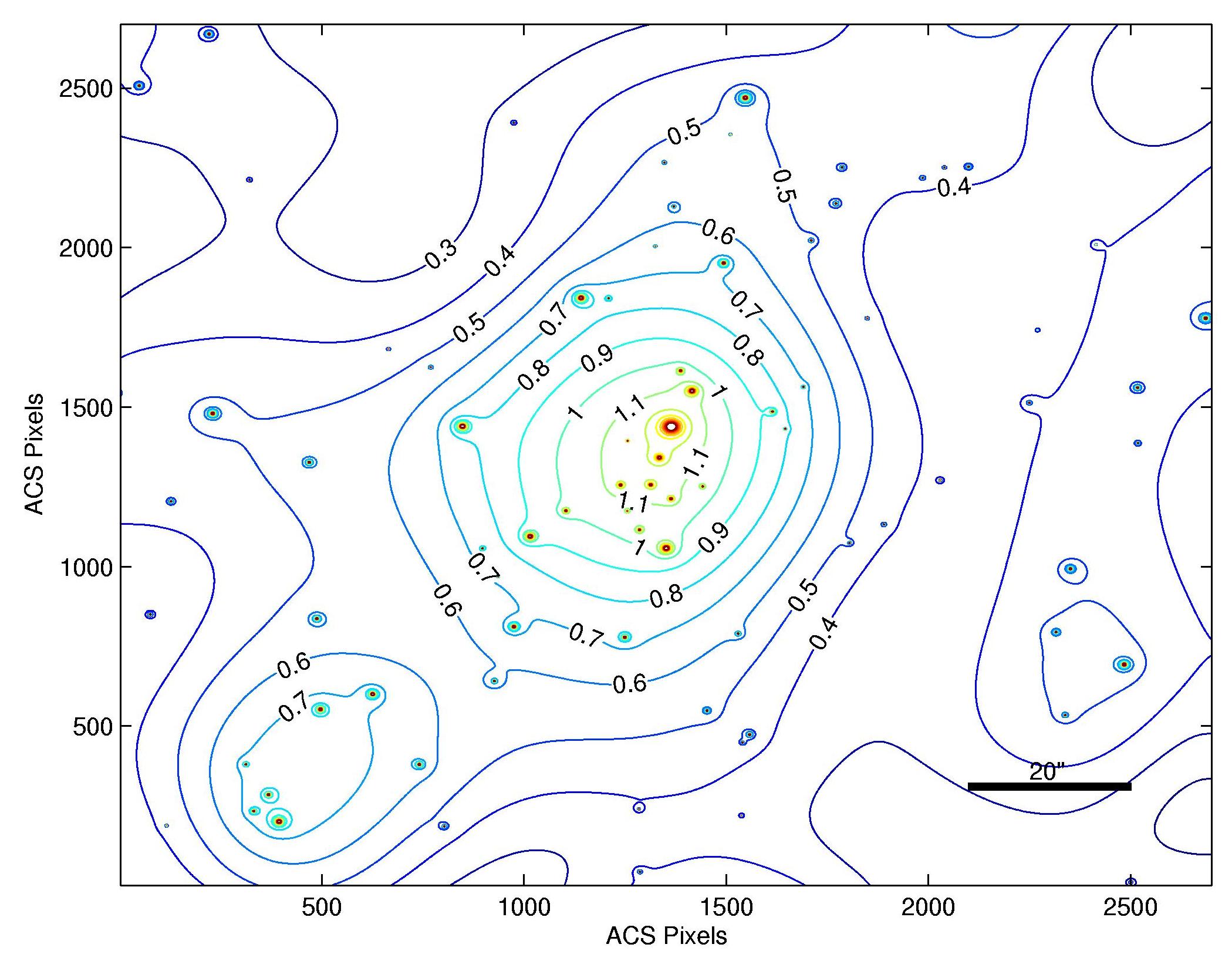}
 \end{center}
\caption{2D surface mass distribution ($\kappa$), in units of the
critical density (for $z_s=4.92$), of MS 1358. Contours are shown in
linear units, derived from our mass model constrained using the many
sets of multiply-lensed images seen in Figure
\ref{curves1358}. Axes are in ACS pixels ($0.05 \arcsec /pixel$), and a 20$\arcsec$ bar is overplotted.}
\label{contoursAdi}
\end{figure}

\begin{figure}
 \begin{center}
   \includegraphics[width=90mm, trim=-8mm 0mm 0mm 0mm,clip]{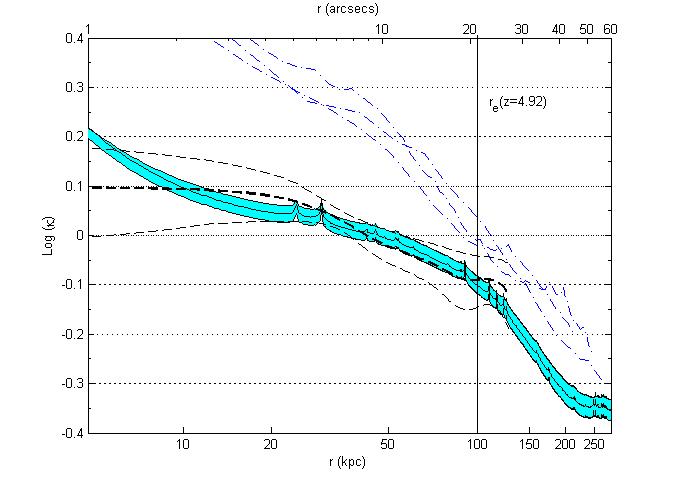}
 \end{center}
\caption{Radial surface mass density ($\kappa$) profile in units of the
critical surface density (for a fiducial redshift of $z_s=4.92$),
derived using the sets of multiple images shown in Figure
\ref{curves1358}. The cyan shaded curve is derived by our modelling
method (Zitrin et al. 2009b), while the dashed black lines are derived
using the non-parametric technique of Liesenborgs et al. (2006). As
can be seen, the profiles agree very well in the region of interest,
and are rather flat. We measure a profile slope of only $d\log \Sigma/d\log \theta\simeq -0.19\pm0.05$ within the effective Einstein radius for system 1 ($r_{E}=21\pm3$; $\simeq100$ kpc; marked with a vertical black line), increasing to $d\log \Sigma/d\log \theta\simeq -0.33\pm0.05$ within twice the critical range, $\simeq200$ kpc. Note that the inner profiles are at
a level close to the critical density ($\pm0.25$) for system 1.
For comparison, overplotted with blue dash-dotted lines are kappa
profiles for three other well-known clusters: Abell 1703, Cl 0024, and
Abell 1689, after correcting for the different lens distances. These were slightly shifted upwards on the y-axis to allow a clean view and to better demonstrate the slope difference of MS 1358 from these typical lensing clusters
which exhibit a common mass slope of $d\log \Sigma/d\log \theta\sim
-0.5$.}
\label{profileAdi}
\end{figure}

In addition to the four previously-known lensed images of the $z_s=4.92$
dropout galaxy (Franx et al. 1997), our modelling technique has
uncovered 19 new multiply-lensed images in the central field of MS
1358, belonging to 7 new systems. We have made use of the location and
photometric redshift information of these images to fully constrain
the mass model. We find that the critical curves for a source at
$z_s=4.92$ (system 1) enclose an area with an effective Einstein
radius of $r_{E}=21\pm3\arcsec$, or $\simeq$100 kpc at the redshift of
the cluster. This critical curve we find (see Figure \ref{curves1358})
encloses a projected mass of $M\simeq 6.1\pm0.8 \times
10^{13}M_{\odot}$. For a source redshift of $z_s\simeq2$ the effective
Einstein radius is $r_{E}=13\pm2\arcsec$, enclosing a projected mass
of $M\simeq 2.7\pm0.2 \times 10^{13}M_{\odot}$ around the BCG. For
general comparison, this is in very good agreement with the Einstein
radius-mass scaling relation for a source at $z_s\simeq2$, found in
Zitrin et al. 2010a (taking into account also the different lens
distances; see Figure 27 therein). The corresponding critical curves
are plotted on the cluster image in Figure \ref{curves1358} along with
the multiply-lensed systems. The resulting mass distribution and its
profile are shown in Figures \ref{contoursAdi} and \ref{profileAdi}.

Note that another small critical curve is formed around the group of
bright cluster galaxies $\sim1\arcmin$ to the south, whose prominence
is hard to determine due to lack of multiple-images at that region. We
note that if this southern clump is somewhat more massive than
presented here, the critical curves may merge with the central main
critical curves to form a larger more elongated critical region. This
however does not seem probable as only few arcs are seen in between,
and thus rule out the existence of such an extended critical curve.

Several mass models have been created in earlier work, describing the
critical curves and mass distribution based only on system~1 ($z=4.92$). Franx
et al. (1997) used isothermal potentials with which they were able to
reconstruct the images of this system. They find an Einstein radius of
$21\arcsec$, similar to our result. Allen (1998) has calculated a total
mass of $8.27 \times 10^{13}M_{\odot}$ with 20\% uncertainty, within a
radius of 121 kpc ($25.5 \arcsec$), in good agreement with our result
considering the Einstein radius difference. Other mass models made for
this cluster based on projected WL profiles and circular
symmetry, derive similar or slightly smaller values (see
Diaferio, Geller \& Rines 2005, e.g., Hoekstra et al. 1998, 2007,
Arabadjis, Bautz \& Garmire 2002, Takahashi \& Chiba 2007).

\begin{figure}
 \begin{center}
  \includegraphics[width=80mm, trim=-8mm 0mm 0mm 0mm,clip]{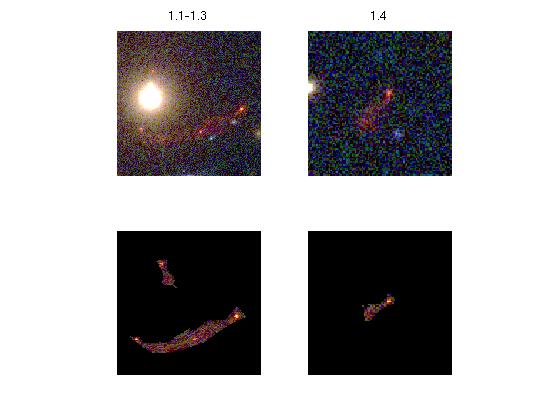}
 \end{center}
\caption{Reproduction of system 1 by our model, by delensing
image 1.1 into the source plane, and then relensing the source plane
pixels onto the image plane. Our model clearly reproduces very
accurately the second half of the arc, 1.2, and the internal structure of
other images in this system including the prominent internal HII regions
within the source.}
\label{rep_sys1}
\end{figure}

\begin{figure}
 \begin{center}
  \includegraphics[width=70mm]{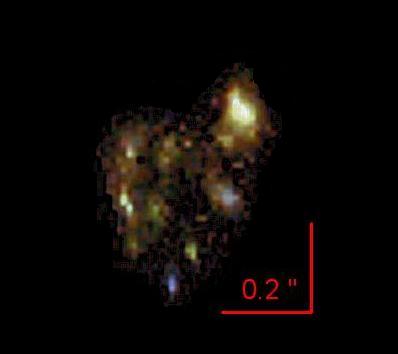}
 \end{center}
\caption{Reproduction of the $z_{s}=4.92$ source galaxy by our model,
using the RIZ images. The image is reproduced by delensing arc 1.1/1.2 into a
high-resolution source-plane chosen to accommodate the
high magnification at its location. Two blue lower-redshift objects got delensed to the source plane in the procedure. Note
that the source shape is very similar to that found in Swinbank et
al. (2009), though due to the magnification difference it is smaller
by a factor of about 4, and seem less internally stretched. The color-coding was slightly modified to obtain a better view of the internal details, and some noise was removed from the edges of the image.}
\label{rep_source1}
\end{figure}

It should be stressed that the multiple-images found here are
accurately reproduced by our model and are not simple identifications by
eye. The parametric method of Zitrin et al. (2009b) has been shown in
many cases to have the predictive power to find multiple images in
clusters. Due to the small number of parameters this model is initially
well-constrained enabling a reliable identification of other
multiple-images in the field, which are then used to fine-tune the
mass model. We now detail each multiply-lensed system in turn, as listed
in Table \ref{systems}:

$System ~1:$ A high-redshift dropout galaxy at $z_s=4.92$. The
four images of this multiply-lensed source were found by Franx et
al. (1997) who also measured its redshift. Two of the multiple-images
of this system (1.1/1.2), form a prominent red fold-arc about
$\sim21\arcsec$ south-west of the BCG, next to another bright cluster
member (see Figure \ref{curves1358}). Image 1.3 is a smaller image on
the other side of that cluster member, and image 1.4 is
$\sim29\arcsec$ east of the BCG. These images are well reproduced by
our model, as seen in Figure \ref{rep_sys1}. According to our model,
the angular area of the source is $\simeq0.1\sq\arcsec$ (see Figure
\ref{rep_source1}), and is therefore magnified about a 100 times in
area, summed over all four images which in total subtend
$\simeq10\sq\arcsec$ in the image-plane (see also \S \ref{magsec}).

$System ~2:$ A long faint arc $\sim21\arcsec$ north of the BCG, next
to a bright cluster member. Our model reproduces this arc accurately
at a redshift of $z_s\simeq3.4$, similar to the photometric redshift
of image 2.1. The photometric redshift of image 2.2 is double peaked,
with peaks around 0.5 and 3.4, the latter in agreement with our
model. In addition, the clearly lensed arc lies away from the BCG and
should have a relative high redshift in order to be lensed, further
strengthening our match. Another similar looking
candidate is seen to the right of image ``c1c'' (see Figure
\ref{curves1358}), but as can be seen the critical curves do not
pass between these images and thus this third candidate is not
probable in the context of our model.

$Systems ~3, 4, and~ 5:$ These three systems consist of typical blue
and white arclets seen in large amounts in lensing clusters. Systems
3-5 follow a similar symmetry, as can been seen in Figure
\ref{curves1358}. One image of each system appears $\sim10\arcsec$
north of the BCG, the second image appears $\sim15\arcsec$ east of the
BCG and the third image $\sim20\arcsec$ to its west. These three
systems have similar typical photometric redshifts of $z\simeq2$ (see
Table \ref{systems}), and due to their close vicinity and similar
appearance were hard to match accurately, especially in the case of
the third images to the west. We therefore note that there can be an
irrelevant mix-up with regards to the third image of each of these
three systems, but since the distances between them are small this
does not affect the resulting mass model. In addition, due to a minor,
northwards offset in the image-plane reproduction of images 4.3 and
5.3, we acknowledge the possibility that there are other
similar-looking spots which may correspond to these systems, but have
a negligible effect on the mass model.

$System ~6:$ Two blue faint images with mirror symmetry lying on two
sides of the critical curve for a source at $z\simeq1.8$ by our
model. Their photometric redshifts are $\simeq2.0\pm0.4$ and
$\simeq1.77^{+0.5}_{-0.4}$, respectively, where the first may be
affected by a nearby galaxy light. Our model predicts the extra small
and faint spec on the other side of the cluster about $\sim20\arcsec$
south-west of the BCG, as identified in the data.

$System ~7:$ Two greenish small images on two sides of a bright
galaxy. These lack photometric redshifts due to the vicinity of the
galaxy, and our model reproduces them accurately at $z\simeq1.8$.

\begin{figure}
 \begin{center}
 \includegraphics[width=80mm, trim=-8mm 0mm 0mm 0mm,clip]{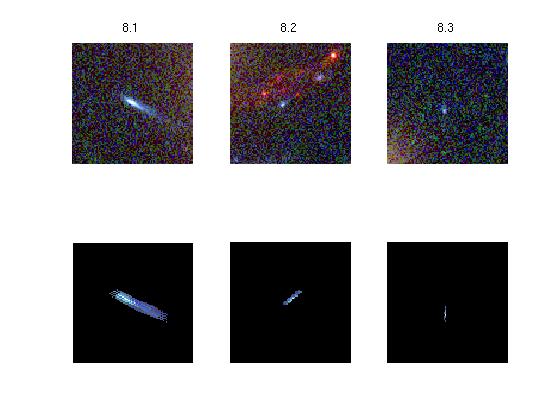}
 \end{center}
\caption{Reproduction of system 8 by our model, by delensing
image 8.1 into the source plane, and then relensing the source plane
pixels onto the image plane. Our model clearly reproduces very
accurately the other images in this system. 
The clear size difference of image 8.1 compared with images 8.2 and
8.3 (zoomed-in here for better view) is another indication of the high
local magnification in that region. }
\label{rep_sys8}
\end{figure}

$System ~8:$ Three blue and white relatively bright images of a
$z\simeq2.1$ galaxy, following similar symmetry as system 1. Our model
reproduces these images very well at this redshift, as seen in Figure
\ref{rep_sys8}. Note in Figure \ref{systems} the highly magnified
region near the location of image 8.1, which explains its larger
appearance relatively to images 8.2 and 8.3.


{\it Other candidate systems and image identification uncertainty:} A
similar symmetry to that systems 2-5 follow, can be expected in the
northern region of the cluster between the BCG and system 2. Moreover,
several images in the west side of the cluster (marked in Figure
\ref{systems}) resemble their corresponding image candidate near
system 2, and have reasonably similar photo-$z$'s. However, according
to our model these candidate systems are not being lensed in full or
in their expected photometric redshifts and we therefore exclude them
from our secure identification.

System ``c1'' consists of two bright blue-white arclets which seem to
have mirror symmetry (``c1a'' and ``c1b''), and an additional similar
looking arclet, ``c1c'', $\sim10\arcsec$ south-west of them. These
have a photometric estimate of $z_s\simeq2$ and as can be seen in
Figure \ref{systems} are north of the critical curve for a such a
source, so that they are not lensed by our model. This may be
artificially overcome if one boosts the mass obtained by the nearby
galaxy, but then system 2 would not correspond to its $z_s\simeq3.4$
photometric redshift. In addition, both this system, and system 2,
have candidates to the west, which are not generated by our model,
since there is no extra apparent mass nor galaxies between these
locations, so the critical curve does not pass between them. One can
significantly boost the nearby galaxies to create a mass lump which
might produce images at that region; however, it should be noted that
such actions are rarely justified and have not been critically
required in any of the many clusters we have analysed to date. We
conclude that despite the similar appearance this system is not
probable in the context of our model.

System ``c2'' is a much more probable candidate consisting of four red
drop-out arclets. Our model reproduces and therefore identifies arcs
``c2c'' and ``c2d'' as the same system in any redshift in the range 3
to 5. Arcs ``c2a'' and ``c2b'' are obtained as part of this system
only in redshifts higher than $\sim4.5$, corresponding to the
photometric redshift of arcs ``c2a'' and ``c2b'', of $\sim4.5$. Still,
we decided not to include it as a secure system since the redshift
agreement is marginal and the distance in the image-plane reproduction
of arcs ``c2a'' and ``c2b'' (from their observed location) is larger
than our typical uncertainty.

Systems ``z1'' and ``z2'' are multiple-images of $z_s\sim7.5$ dropout
candidates, claimed as possibilities by Richard et al. (2008). In
Figure \ref{curves1358} we mark their \emph{approximate} locations generated by
our model (see Figure 12 in Richard et al. 2008). We did not use these
candidate images to constrain the model, but verified that they are plausible
high-$z$ objects in the context of our model (for $z_{cl}=0.33$ the
lensing-distance of sources at $z_s\sim7.5$ is only $\simeq2\%$ bigger
than the lensing distance of the $z_s=4.92$ galaxy, so that redshifts
throughout that range are plausible in that context).

\subsection{Mass Profile and High Magnification}\label{magsec}

We mentioned in the preceding sections that the profile can only be
accurately constrained by incorporating the cosmological
redshift-distance relation, i.e., the lensing distance of each system
based on the measured spectroscopic or photometric redshifts.  In so
doing we normalise our mass model to system 1, so that the normalised
scaling factor, $f(d_{ls}/d_{s})$, is equal to 1. We then make use of
the lower-$z$ systems, in particular the $z\simeq2$ systems (number
3,4,5 and 8) and the $z\simeq3.4$ system (number 2), whose photometric
redshifts we find most credible to constrain the profile.

\begin{figure}
 \begin{center}
   \includegraphics[width=95mm, trim=-8mm 0mm 0mm 0mm,clip]{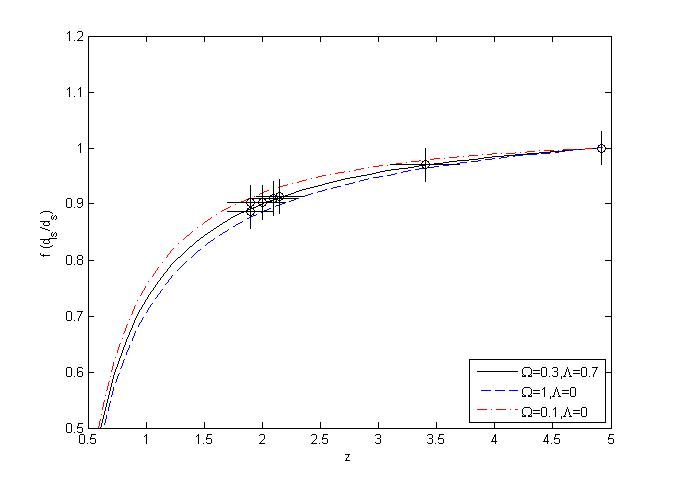}
 \end{center}
\caption{Growth of the scaling factor $f(d_{ls}/d_{s})$ as a function of
redshift, normalised so f=1 at $z=4.92$. Plotted lines are the
expected ratio from the chosen specified cosmological model. The
circles correspond to the multiple-image systems reproduced by the
parametric mass model, versus their real photometric redshift. The
data follow very well the relation predicted by the standard
cosmological model (mean deviation of only $\Delta_{f}< 0.01$, and
$\chi^{2}\simeq0.1$ for this fit). We take an average photo-z for
each system, where there is an estimation difference between the same
images of the same system.}
\label{dlsds}
\end{figure}

We examine how well the cosmological relation is reproduced by our
model, accounting for all systems with photo-$z$'s, as shown in Figure
\ref{dlsds}. Clearly the redshifts of these systems verify very well
that the predicted deflection of the best fitting model at the
redshift of each of these systems, lies accurately along the expected
cosmological relation, with a mean deviation of only $\Delta_{f}<
0.01$ (see Figure \ref{dlsds}), and $\chi^{2}\simeq0.1$ for the best
model, considerably strengthening the plausibility of our approach to
modelling in general.

We find that unlike other well-known lensing clusters such as Abell
1689, Abell 1703 or Cl0024+17, which all have a typical profile slope
of $d\log \Sigma/d\log \theta\simeq -0.5$ (see Broadhurst et
al. 2005a, Zitrin et al. 2009b, 2010b), MS 1358 has a relatively
shallow profile slope of only $d\log \Sigma/d\log \theta\simeq
-0.19\pm0.05$ in its main critical region of
$\sim$[3\arcsec,21\arcsec], increasing to $d\log \Sigma/d\log \theta\simeq
-0.33\pm0.05$ in twice the critical region $\sim$[3\arcsec,42\arcsec] ($\simeq200$ kpc), as can be seen in Figure
\ref{profileAdi}. We compare our profile with the profile derived by
the assumption-free method of Liesenborgs et al. (2006), overplotted
as well in Figure \ref{profileAdi}. As can be seen the profiles are
similarly shallow and in good agreement with each other. This may be
expected, since the profile must be adjusted to maintain the
lensing-distance ratio for each lensed system, manifesting the
importance of the images found in this work for constraining the mass
profile of this cluster.

The shallow profile results in high magnification, as kappa is close
to unity over a wide central area and thus the magnification is
boosted. This phenomenon we have revealed recently in another even
shallower profile cluster, MACS 1149.5+2223 (Zitrin \& Broadhurst
2009) which hugely magnifies a distant spiral galaxy by a factor of
$\sim200\times$ creating the largest lensed images known, as the slope is
even shallower in this case related to the unrelaxed nature of that
cluster. Another work based on spectroscopic redshifts claims a
somewhat more moderate slope of up to $d\log \Sigma/d\log \theta\simeq
-0.3$ for MACS 1149.5+2223 (Smith et al. 2009) but still significantly
lower than typically found in other rich lensing clusters. In MS 1358
the high magnification is inferred by the distribution of
multiple-images and their redshifts, and can account for the
large bright images of system 1, at $z_s=4.92$ (see
Figure
\ref{curves1358}). We find that the source galaxy is magnified
$\sim80\times$ into arcs 1.1/1.2, and about $5\times-15\times$ into each of arcs 1.3 and 1.4, so that the source is magnified in total
about a 100 times, and is therefore one of the most highly-magnified
distant objects known.

The magnification values are hard to determine precisely since the
magnification is a very sensitive function of the local mass gradient
so that in practice shallower models yield much higher
magnifications. For example, Franx et al. (1997), have used isothermal
potentials to find a magnification of $\times5-\times11$ at the
eastern part of the bigger arc (image 1.2 here), and Swinbank et
al. (2009; see also Richard et al. 2008) have found a revised
``luminosity-weighted'' magnification of $\simeq12.5$, while we find
in that region an average magnification of $\sim\times25-\times50$. It
is well understood now by most practitioners that the mass profile
cannot be appropriately constrained when too few multiple-images are
used, or when they span only a narrow redshift range. Here, our
identification of many multiple-images spread over a wide range of
redshifts allows for the first time an accurate determination of the
profile slope of MS 1358, revealing the rather shallow central mass profile.

\subsection{Luminosity of the $z=4.92$ Source}

Due to the large difference in magnification with models published
earlier (Franx et al. 1997, Swinbank et al. 2009), and due to the
importance of the object at $z_s=4.92$, we calculate our corrections
to the original $z=4.92$ source luminosity. Firstly, we simply divide
the bolometric luminosity published by Franx et al.,
$\sim3\times10^{11} L_{\odot}$, by the approximate magnification ratio of the two
mass models, to obtain a corrected source bolometric luminosity of
$\sim6\times10^{10} L_{\odot}$. Independently, we find that the source
AB magnitude is I$_{814,AB}=25.67$ and I$_{775,AB}=25.76$ (after
accounting for the magnification), significantly fainter than
I$_{814,AB}=24.0$ or I$_{775,AB}=24.94$, as published by Franx et
al. and Swinbank et al., respectively. We then use eq. 2
of Bouwens et al. (2009a) to convert the corrected bolometric luminosity into an
absolute magnitude in the rest-frame UV. This relation is known to
hold from $z\sim2.5$ to at least $z\sim4$, and is expected to be
approximately valid also at $z=4.92$. Following our model, the absolute magnitude
of the source is $M^{*}_{UV, AB}\simeq-19.5$, $\sim4$ times a more
common value for objects at $z\sim5$ according to up-to-date high-$z$
luminosity functions (e.g., Bouwens et al. 2007, van der Burg,
Hildebrandt \& Erben 2010).

\subsection{Mass-To-Light Ratio}

We calculate here the central Mass-To-Light ratio using the Luminous Red
Galaxies (LRG) template described in Ben\'itez et al. (2009) to
convert fluxes into luminosities, and measure an $M/L_{B}$ ratio of
$\simeq160\pm30$ $(M/L)_{\odot}$ within the critical curves for $z_{s}=4.92$ (see Figure \ref{curves1358}). We compare this value to
the M/L versus Einstein radius relation found in Zitrin et al. (2010a)
for a sample of 12 clusters. After correcting for the difference in
the lens redshift, this value tightly follows the relation they
presented (see Figure 28 therein), which predicts a similar ratio.

We may also examine in detail the M/L of the luminous elliptical galaxy
which splits the images of the $z_{s}=4.92$ galaxy arcs 1.1/1.2 and 1.3 (visible in
Figure \ref{curves1358}; RA=13:59:49.47 DEC=
+62:30:47.55). According to our model this cluster member has a mass
of $1.45^{+0.30}_{-0.40} \times 10^{11} M_{\odot}$ within a radius of
$3\arcsec$ ($\sim14$ kpc), after subtracting the cluster smooth DM
component. We compare this result with the mass inferred by the virial
theorem through the velocity dispersion, and in the same aperture. For
example, Kelson et al. (2000a,b, 2006) measured a velocity dispersion
of $\sigma\sim 220$ km/s for this galaxy, which corresponds to $\sim1.6
\times 10^{11} M_{\odot}$ within $3\arcsec$, in fair agreement with
our result (we note however that a smaller galaxy may be hidden in
this bright elliptical light, slightly affecting the comparison). The
light enclosed within $3\arcsec$ is $\sim6\times10^{10} L_{\odot}$,
which yields a small $M/L_{B}$ ratio of $\sim3$ $(M/L)_{\odot}$, close to the purely stellar value
indicating no significant DM out to $\sim14$ kpc. The
effective radius of this galaxy was found to be $\simeq1\arcsec$ (4.75
kpc; Kelson et al. 2000a,b, 2006) so that in practice we measure in a
radius bigger three times than the effective radius, where the DM
should be dominant. However, a lower $M/L_{B}$ ratio is not surprising, because we subtracted the cluster DM component from our mass
measurement, and the halo of this galaxy should already be stripped by the significant tidal forces in the central
cluster region (if not earlier during the hierarchical evolution of the
cluster through mergers of subclumps). It is known that about 70\%
of the DM is stripped from central galaxies also in simulations of massive clusters (e.g., Nagai \& Kravtsov 2005).

\section{Summary}

In this work we have presented a detailed lensing analysis of the
galaxy cluster MS 1358 in HST/ACS images. Our well-established
modelling method (Broadhurst et al. 2005a, Zitrin et al. 2009b, Zitrin et al. 2010a,b) has
revealed the rather shallow mass distribution of the central region by
uncovering 19 multiply-lensed images which eluded previous detection,
so that in total 23 images of 8 different sources were used to fully
constrain the fit. Though more lensed candidates might be found in
this field with further careful effort, our minimalistic approach to
lensing involves only 6 free parameters so that the resulting model is
clearly fully constrained by these multiple images. In addition we
have uncovered several other lensed candidates, for which it would be
interesting to obtain spectroscopic redshifts (as well as to the
systems identification presented here) in order to further establish the
results of this work.

The photometric redshifts of the newly found arcs throughout the
central region enable the determination of the inner mass profile of
MS 1358, through the cosmological lensing-distance ratio, and imply a
shallow mass distribution manifested also in our modelling iterations
so that models spanning almost the full parameter space yield similar
shallow mass profiles with typical slopes of $d\log \Sigma/d\log
\theta\simeq -0.3\pm0.1$ ($r<200$ kpc). We further tested our results with a
non-parametric adaptive-grid method (Liesenborgs et al. 2006, 2007,
2009), which yields a similarly shallow profile hence strengthening
the conclusions of this work. A shallow mass distribution translates
into a boosted magnification in the central region so that the
$z=4.92$ source galaxy is magnified in total $\sim100$ times, and is
therefore one of the more highly-magnified, distant objects known.

Our magnification values for this system are higher by a factor of
$\sim5$ than previous estimates (e.g., Franx et al. 1997, Swinbank et al. 2009) based on models constrained only by the
$z=4.92$ system. Hence the source size, bolometric luminosity, and star-formation rate are correspondingly
$\sim5\times$ smaller, more typical of faint field galaxies
based on current high-$z$ luminosity functions (e.g., Bouwens et
al. 2007, van der Burg, Hildebrandt \& Erben 2010), and with a physical
scale of a dwarf galaxy of radius $\simeq 1$ ~kpc. A detailed image of the
interior morphology of the source is generated by delensing the most
magnified image (1.1/1.2), resulting in a high effective resolution of only
50 pc, thanks to the high magnification and to the declining angular
diameter distance above $z\simeq1.5$ for the standard cosmology, so that this
image apparently represents the best resolved object known at high redshift.
%

\section*{acknowledgments}

This research is being supported by the Israel Science Foundation, and encouraged further by the TAU school of physics and
astronomy scholarship for research excellency. AZ acknowledges Eran
Ofek and Salman Rogers for their publicly available Matlab scripts. We
thank Daniel Kelson for useful information. ACS was developed under
NASA contract NAS 5-32865. Results are based on observations made with
the NASA/ESA Hubble Space Telescope, obtained from the data archive at
the Space Telescope Science Institute. STScI is operated by the
Association of Universities for Research in Astronomy, Inc. under NASA
contract NAS 5-26555.

\bsp
\label{lastpage}

\end{document}